\documentclass[10pt]{iopart}

\usepackage{amssymb}

\newcommand{\bC} {\mathbb{C}}
\newcommand{\bN} {\mathbb{N}}
\newcommand{\bR} {\mathbb{R}}
\newcommand{\cH} {\mathcal{H}}
\newcommand{\cP} {\mathcal{P}}
\newcommand{\cR} {\mathcal{R}}
\newcommand{\hH} {\hat{\cH}}

\newcommand{\fexp}  [1] {\exp \left( #1 \right)}
\newcommand{\fabs}  [1] {\left| #1 \right|}
\newcommand{\fabsq} [1] {\left| #1 \right|^2}
\newcommand{\fsqrt} [1] {\sqrt{#1} \,}
\newcommand{\fnorm} [1] {\left\| #1 \right\|}
\newcommand{\fnormq}[1] {{\left\| #1 \right\|}^2}

\newcommand{\ket}[1] {\,|#1\!>\,}
\newcommand{\bra}[1] {\,<\!#1|\,}

\newcommand{\Theorref}[1]{Theorem~\ref{#1}}
\newcommand{\eqref}[1]{(\ref{#1})}

\newcounter{theorem}
\newenvironment{theorem}{\vspace{0.2cm}\stepcounter{theorem}
{\parindent0cm \bf Theorem \arabic{theorem}.}\begin{it}}{\end{it}\vspace{0.2cm}}

\newenvironment{definition}{\vspace{0.2cm}
{\parindent0cm \bf Definition.}\begin{it}}{\end{it}\vspace{0.2cm}}

\newenvironment{lemma}{\vspace{0.2cm}
{\parindent0cm \bf Lemma.}\begin{it}}{\end{it}\vspace{0.2cm}}

\newenvironment{proof}{{\parindent0cm \bf Proof.}
}{\hfill\fbox{}\vspace{0.2cm}}

\begin{document}

% ===============================================================================
% Title 
% ===============================================================================

\title{A Relativistic Extension of Event-Enhanced Quantum Theory}
\author{Andreas Ruschhaupt}
\address{Faculty of Physics, University of Bielefeld,
Universit\"atsstr. 25,\\
D-33615 Bielefeld, Germany}
\ead{rushha@physik.uni-bielefeld.de}

\begin{abstract}
In this paper, we develop a formalism describing in a relativistic way
a system which consists of a classical and a quantum part being
coupled. The formalism models one particle with spin $\frac{1}{2}$ and
it is a possible relativistic extension of the Event-Enhanced Quantum Theory.
We postulate a covariant algorithm
which plays the role of the standard reduction postulate in
non-relativistic quantum mechanics. Furthermore, we present an
algorithm to simulate detections of the particle.
\end{abstract}


% ===============================================================================
% Section 1
% Introduction
% ===============================================================================

\section{Introduction}
Seeking to bridge the conceptual frameworks of classical and
quantum theory, Blanchard and Jadczyk \cite{blanchard.1993a,
blanchard.1995c,blanchard.1995d} have proposed an extension of standard
(non-relativistic) quantum mechanics called Event-Enhanced Quantum Theory (EEQT). 
Its main idea is to view the total system as consisting of a classical
and a quantum part which are coupled. The pure states
of the quantum part are wave functions which
are not directly observable, whereas the pure states of the classical part can 
be observed without disturbing them. Changes of the classical pure
states are called events. Events are discrete and irreversible.
A review about applications of EEQT is for example
\cite{blanchard.2000}.

Trying to define states and a reduction postulate in a relativistic theory can lead
to paradoxes and logical difficulties (for example see Y. Aharonov and D.Z. Albert
\cite{aharonov.1984}).

One possibility to avoid some of these difficulties is to consider the wave function
for relativistic particle not as a function on the space-time continuum but as a
function on the set of flat, space-like hypersurfaces in Minkowski space (for
example see the papers by Breuer and Petruccione \cite{breuer.1998a,
breuer.1998b, breuer.1999b}).

Another possibility is the introduction of a supplementary,
intrinsic time, the proper time $\tau$. The proper time $\tau$ is independent of the
reference frame. It plays the role of (absolute) time in
non-relativistic quantum mechanics. The idea of a proper time was
first used physically by Horwitz and Piron
\cite{horwitz.1973} and later in a lot of other approaches (a review
with more references is for example written by Fanchi \cite{fanchi.1993}). 

Blanchard and Jadczyk have also introduced a relativistic
version of EEQT \cite{blanchard.1996b} using the idea of a proper time
and an indefinite scalar product.

The main aim of this paper is to present an alternative relativistic version of
EEQT which uses a positive-definite scalar product.
The theory will describe a single spin $\frac{1}{2}$ particle with mass
$m$ in a relativistic way and should be useful in situations in which one can neglect
pair-creation and pair-annihilation.
As in the relativistic extension introduced by Blanchard and Jadczyk
\cite{blanchard.1996b}, we postulate an additional parameter, called
proper time $\tau$. 
The total system consists of a classical
and a quantum part. Therefore, at a given proper time $\tau$, the (pure) state of
the total system is a pair $(\omega_\tau,\Psi_\tau)$. $\omega_\tau$ 
is the state of the classical part and $\Psi_\tau$ is the state of the quantum part.

We assume that a (pure) state  $\omega_\tau$ of the classical part is a number:
$\omega_\tau \in \bN_0=\{0,1,2,..\}$. Again, a change of the classical
(pure) state is called an ``event''.

The (pure) states of the quantum part shall be (heuristically spoken) solutions
$\Psi : \bR \times \bR^3 \to \bC^4$ of
the Dirac-equation $\left(\rmi \gamma^\mu \partial_\mu - \frac{e}{c\hbar} \gamma^\mu A_\mu -
    \frac{mc}{\hbar} \right) \Psi (x) = 0$. 
An interesting property of a quantum state is that it is uniquely given by its
values on a spacelike hyperplane.
Moreover, it will be possible to introduce a positive-definite scalar
product between two quantum states. In the second section of this paper, we will
present the definition of (pure) states of the quantum part and their
properties in a more mathematical way.

We define in the third section how the system state changes if we change the reference
frame or ``charge conjugate'' the system.   

In the forth section, we postulate a covariant algorithm for simulating ideal measurements of
infinitesimally small duration. It plays the role of the standard
reduction postulate in non-relativistic quantum mechanics.

An algorithm for simulating detections of the particle is
presented and examined in the fifth section.

In the last section, we summarize the properties of our formalism.

In a future paper, we will examine applications of our algorithm for
simulating detections. A first application can be
found in \cite{ruschhaupt.1999}.

% ===============================================================================
% Section 2
% Pure states of the quantum part
% ===============================================================================

\section{Pure states of the quantum part}

We want to define a (pure) state of the quantum part of the total
system. It describes the state of a single particle with spin
$\frac{1}{2}$ and mass $m$. 

Let $\cP  =  \left\{ (y,\vec{\alpha},\vec{\varphi}) : y \in \bR^4, \vec{\alpha} \in
  \bR^3 , \fabs{\vec{\alpha}} < 1, \vec{\varphi} \in \bR^3,
  \fabs{\vec{\varphi}} < \pi \right\}$ and we define  with
  $\lambda\equiv((y^0,\vec{y}),\vec{\alpha},\vec{\varphi})~\in~\cP$:
\begin{eqnarray*}
\begin{array}{rcll}
  \sigma_\lambda (\vec{u}) & = & \left( y^0 +
  \vec{\alpha} \cdot \hat{R}_{\vec{\varphi}}\vec{u}, \; \vec{y} +
  \hat{R}_{\vec{\varphi}}\vec{u} \right) & \forall \vec{u} \in \bR^3 \\
  <f|g>_\lambda & = & \int \rmd\vec{u} \,
  f^+(\vec{u}) \left( 1 - \gamma^0 \vec{\gamma}\vec{\alpha} \right)
  g(\vec{u}) & \forall f,g \in L_2(\bR^3)^4 \\
  \fnorm{f}_\lambda & = & \fsqrt{<f|f>_\lambda} & \forall f \in L_2(\bR^3)^4
\end{array}
\end{eqnarray*}
with $\bar{\bC} = \bC \cup \{\pm\infty + \rmi\bR\} \cup
\{\bR\pm\rmi\infty\} \cup \{\pm\infty \pm\rmi\infty\}$
and $L_2 (\bR^3)^4 = \left\{ f : \bR^3 \to \bar{\bC}^4 : 
\int \rmd x \fabsq{f(x)} < \infty \right\}$.
$\gamma^\mu=(\gamma^0,\vec{\gamma})$ are the Dirac matrices and
$\hat{R}_{\vec{\varphi}} \in SO(3)$ is the
rotation of the angle $\fabs{\vec{\varphi}}$ around the vector
$\vec{\varphi}/\fabs{\vec{\varphi}}$ (the sense of rotation is
determined by the right-hand rule).
We continue with the following definition:

%
% ------------------------------ Definition -------------------------------------
%

\begin{definition}
$\Psi \in \hH$ if and only if the following conditions are satisfied
\begin{eqnarray}
\fl (i)\; &
\Psi : \bR \times \bR^3 \to \bC^4, \Psi \; \mbox{continuous differentiable}\\
\fl (ii)\; &
    \left(\rmi \gamma^\mu \partial_\mu - \frac{e}{c\hbar} \gamma^\mu A_\mu -
      \frac{mc}{\hbar} \right) \Psi (x) = 0
      \label{sec2_dirac_eq} \\
\fl (iii)\; &
\fnorm{\Psi \circ \sigma_\lambda}_\lambda <
\infty \quad \mbox{for all} \quad \lambda \in \cP
\label{sec2_def_iii}\\
\fl (iv)\; &
\lim_{\fabs{\vec{u}} \to \infty} \fabs{\vec{u}}^3
\fabsq{\Psi \circ \sigma_\lambda (\vec{u})} = 0 \quad \mbox{for all}
\quad \lambda \in \cP
\label{sec2_def_iv}
\end{eqnarray}
$A_\mu : \bR^4 \to \bR^4$ is the external electromagnetic potential.
\end{definition}

$\hH$ is a vector space. Now we want to define a scalar product for
all $\Psi \in \hH$. The next theorem is very important for achieving this task.

%
% ------------------------------ Theorem 1 --------------------------------------
%

\begin{theorem} Let $\Psi_A, \Psi_B \in \hH$, let $ j_{AB}^\mu := \Psi^+_A
  \gamma^0 \gamma^\mu \Psi_B$, the quantity
\begin{eqnarray*}
 <\Psi_A \circ \sigma_\lambda | \Psi_B \circ \sigma_\lambda>_\lambda
 \, \equiv \;
\int_{\sigma_\lambda} j_{AB}^\mu \rmd f_\mu
\end{eqnarray*}
exists for all $\lambda=((y^0, \vec{y}),\vec{\alpha},\vec{\varphi})
\in \cP$ and is independent of 
$\lambda$. $\rmd f_\mu~\equiv~(1,-\vec{\alpha})~\rmd\vec{u}$ denotes the differential
``surface element'' of $\sigma_\lambda$.
\label{theorem_1}
\end{theorem}

%
% ------------------------------ Proof of Theorem 1 -----------------------------
%

\begin{proof}
(i) existence:
This follows from the fact that $\Psi_A \circ \sigma_\lambda,  \Psi_B \circ
\sigma_\lambda \in L_2 (\bR^3)^4$ (see \eqref{sec2_def_iii}).
\\
(ii) independence:
We get $\partial_\mu j_{AB}^\mu = 0$ by a simple calculation.
The integral is clearly independent of $\vec{\varphi}$ and $\vec{y}$. Therefore, we can assume
$\vec{\varphi} = 0$ and $\vec{y} = 0$. Let
$\sigma_1 = \sigma_{((y_1^0,\vec{0}),\vec{\alpha}_1,0)}$ and 
$\sigma_2 = \sigma_{((y_2^0,\vec{0}),\vec{\alpha_2}, 0)}$
be two hyperplanes. Let 
$\hat{x} (\varphi, \Theta) = \left( \cos \varphi \sin \Theta,
\sin \varphi \sin \Theta, \cos \Theta \right)$. 
\\
(a) case $\vec{\alpha_1} = \vec{\alpha_2} =: \vec{\alpha}$: Let
\begin{eqnarray*}
F_1 (R) & = & \{ \sigma_1 (\vec{u}) : \fabs{\vec{u}} \le R \} \\
F_2 (R) & = & \{ \sigma_2 (\vec{u}) : \fabs{\vec{u}} \le R \} \\
s_R (\nu, \varphi, \Theta) & = & \left( 
y_1^0 + \nu (y_2^0 - y_1^0) + R \cdot \vec{\alpha} \hat{x} (\varphi,\Theta),
R \cdot \hat{x} (\varphi,\Theta) \right)\\
S(R) & = & \{ s_R (\nu, \varphi, \Theta) : 0 \le \nu \le 1, 0 \le \varphi < 2\pi,
0 \le \Theta < \pi \}
\end{eqnarray*}
Let $V(R)$ be the volume bounded by $F_1(R), F_2 (R)$, and $S (R)$. 
The differential ``surface element'' of $S(R)$ is 
$\rmd S_\mu = R^2 W_\mu (\nu, \varphi, \alpha) \, d\nu \, d\varphi \,
d\Theta$.
The function $W_\mu$ need not to be explicitly calculated, because
it is enough to know that $W_\mu$ does not depend on $R$.
We get by Gauss theorem (with $j_{AB}^\mu
(R,\nu,\varphi,\Theta) \equiv j_{AB}^\mu \circ s_R (\nu,\varphi,\Theta)$)
\begin{eqnarray*}
\fl -\int_{\sigma_1} j_{AB}^\mu \rmd f_\mu + \int_{\sigma_2} j_{AB}^\mu \rmd f_\mu =
- \lim_{R\to\infty} \int_{F_1(R)} j_{AB}^\mu \rmd f_\mu + \lim_{R\to\infty}
\int_{F_2(R)} j_{AB}^\mu \rmd f_\mu \\
\lo= \lim_{R\to\infty} \int_{V(R)} \partial_\mu  j_{AB}^\mu \rmd^4x -
\lim_{R\to\infty} \int_{S(R)} j_{AB}^\mu \rmd S_\mu \\
\lo= -\lim_{R\to\infty} \int_{S(R)} j_{AB}^\mu \rmd S_\mu \\
\lo= -\lim_{R\to\infty} \int \rmd\nu \int \rmd\varphi \int \rmd\Theta \, R^2 j_{AB}^\mu
(R,\nu,\varphi,\Theta) W_\mu (\nu, \varphi, \alpha) \\
\lo= -\int \rmd\nu \int \rmd\varphi \int \rmd\Theta \lim_{R\to\infty} (R^2 j_{AB}^\mu
(R,\nu,\varphi,\Theta)) W_\mu (\nu, \varphi, \alpha) = 0
\end{eqnarray*}
because
\begin{eqnarray*}
\fl R^2 \fabs{j_{AB}^\mu (R,\nu,\varphi,\Theta)} = R^2 \fabs{\Psi_A^+ \gamma^0
  \gamma^\mu \Psi_B \circ s_R (\nu,\varphi,\Theta)}\\
\lo\le \frac{const}{2} \left( R^2 \fabsq{\Psi_A \circ s_R (\nu,\varphi,\Theta)} +
  R^2 \fabsq{\Psi_B \circ s_R (\nu,\varphi,\Theta)} \right)
  \stackrel{R\to\infty}{\longrightarrow} 0
\end{eqnarray*}
uniformly in $\varphi,\Theta$ (see \eqref{sec2_def_iv}) and $\nu$ (because
$j_{AB}^\mu$ is continuous).
\\
(b) case $\vec{\alpha_1} \neq \vec{\alpha_2}$:
Because of case (a), we can assume $y^0_1 = y^0_2 = 0$. Let 
$\vec{\alpha}(\nu)$ be chosen in such a way that 
$\vec{\alpha}(\nu)$ is continuous, $\vec{\alpha}(0) =
\vec{\alpha_1}$, $\vec{\alpha}(1) = \vec{\alpha_2}$ and
$\fabs{\vec{\alpha}(\nu)} < 1 \, \forall \nu\in[0..1]$. Let
\begin{eqnarray*}
F_1 (R) & = & \{ \sigma_1 (\vec{u}) : \fabs{\vec{u}} \le R \} \\
F_2 (R) & = & \{ \sigma_2 (\vec{u}) : \fabs{\vec{u}} \le R \} \\
s_R (\nu, \varphi, \Theta) & = & 
\left( R \cdot \vec{\alpha}(\nu) \hat{x} (\varphi,\Theta),
R \cdot \hat{x} (\varphi,\Theta) \right)\\
S(R) & = & \{ s_R (\nu, \varphi, \Theta) : 0 \le \nu \le 1, 0 \le \varphi < 2\pi,
0 \le \Theta < \pi \}
\end{eqnarray*}
Again, $V(R)$ should be the volume bounded by $F_1(R), F_2 (R)$, and $S (R)$.
The differential ``surface element'' of $S(R)$ is
$\rmd S_\mu = R^3 \tilde{W}_\mu (\nu, \varphi, \alpha) \, d\nu \, d\varphi \, d\Theta$
(note the factor $R^3$ instead of $R^2$ in case (a)!). 
Analog to case (a), it follows
\begin{eqnarray*}
-\int_{\sigma_1} j_{AB}^\mu \rmd f_\mu + \int_{\sigma_2} j_{AB}^\mu \rmd f_\mu & = & 0
\end{eqnarray*}
because
$\fabs{R^3 j_{AB}^\mu (R,\nu,\varphi,\Theta)}
\stackrel{R\to\infty}{\longrightarrow} 0$ uniformly in $\nu,\varphi,\Theta$.
\end{proof}

Now we are able to introduce a scalar product between elements of $\hH$:

%
% ------------------------------ Definition -------------------------------------
%

\begin{definition} We introduce a scalar product between $\Psi_A, \Psi_B \in
  \hH$:
\begin{eqnarray*}
  <\Psi_A | \Psi_B>_{\hH} & := & 
  <\Psi_A \circ \sigma_\lambda | \Psi_B \circ \sigma_\lambda >_\lambda\\
  \fnorm {\Psi_A}_{\hH} & := & \fsqrt{<\Psi_A | \Psi_A>_{\hH}}
\end{eqnarray*}
with $\lambda \in \cP$ arbitrary.
\end{definition}

$<.|.>_{\hH}$ is a sesquilinear form. It is clear that
$<\Psi|\Psi>_{\hH} \, \ge 0 \; \forall \Psi \in \hH$ because the
eigenvalues of $(1 - \gamma^0 \vec{\gamma} 
\vec{\alpha})$ are $1 + \fabs{\vec{\alpha}} > 0$ and $1 -
\fabs{\vec{\alpha}} > 0$.

The independence of the scalar product from the parameters $\lambda \equiv
(y,\vec{\alpha}, \vec{\varphi})$ ``expresses'' the independence of the reference frame. 
Note that the number of ``free parameters'' is
ten and equals the number of parameters of a Poincar\'e-transformation.

An element $\Psi \in \hH$ is uniquely given by its values on a hyperplane
$\sigma_\lambda$. This fact results indeed from the next theorem.

%
% ------------------------------ Theorem 2 --------------------------------------
%

\begin{theorem} Let $\mu=(y,\vec{\alpha},\vec{\varphi}) \in \cP$ arbitrary,
  let $\Psi_1, \Psi_2 \in \hH$ with $\Psi_1 \circ \sigma_\mu = \Psi_2 \circ \sigma_\mu$
then it follows $\Psi_1 = \Psi_2$.
\label{theorem_2}
\end{theorem}

%
% ------------------------------ Proof of Theorem 2 -----------------------------
%

\begin{proof}
Let $\Psi := \Psi_1 - \Psi_2$, we get $\Psi \circ
\sigma_\mu (\vec{u}) = 0 \; \forall \vec{u}$ and therefore
$\fnorm{\Psi \circ \sigma_\mu}_\mu = 0$.
We assume $\Psi_1 \neq \Psi_2$, so it exist $z = (z^0,\vec{z}) \in \bR^4$ with
$\Psi (z) = \Psi_1(z) - \Psi_2(z) \neq 0$. Because $\Psi$ is continuous there
must be a neighborhood of $z$ with $\Psi (x) \neq 0$. So it exists $\epsilon >
0$ with $\Psi \circ \sigma_{(z,\vec{\alpha},\vec{\varphi})} (\vec{u}) \neq 0$ for all $\vec{u}$
with $\fabs{\vec{u}} < \epsilon$ (because $z = \sigma_{(z,\vec{\alpha},\vec{\varphi})} (0)$).
It follows that
$\fnorm{\Psi \circ
\sigma_{(z,\vec{\alpha},\vec{\varphi})}}_{\sigma_{(z,\vec{\alpha},\vec{\varphi})}}
> 0$.
But we get
\begin{eqnarray*}
0 = \fnorm{\Psi \circ
 \sigma_{(y,\vec{\alpha},\vec{\varphi})}}_{\sigma_{(y,\vec{\alpha},\vec{\varphi})}}
 \stackrel{\Theorref{theorem_1}}{=} \fnorm{\Psi \circ 
 \sigma_{(z,\vec{\alpha},\vec{\varphi})}}_{\sigma_{(z,\vec{\alpha},\vec{\varphi})}}
 \neq 0
\end{eqnarray*}
The assumption that $\Psi_1 \neq \Psi_2$ is wrong and it implies that $\Psi_1 =
\Psi_2$.
\end{proof}

%
% ------------------------------ Theorem 3 --------------------------------------
%

\begin{theorem} 
$(\hH, <.|.>_{\hH})$ is a pre-Hilbert space.
\end{theorem}

%
% ------------------------------ Proof of Theorem 3 -----------------------------
%

\begin{proof}
It is only left to proof that $<\Psi|\Psi>_{\hH} = 0$ provides $\Psi = 0$. We
assume
\begin{eqnarray*}
0 \, = \, <\Psi|\Psi>_{\hH} \, = \int \rmd\vec{u} \fabsq{\Psi (0,\vec{u})} 
= \, <\Psi|\Psi>_{L_2 (\bR^3)^4}
\end{eqnarray*}
It results that $\Psi (0,\vec{u}) = 0 \, \forall \vec{u}$, because
$<.|.>_{L_2(\bR^3)^4}$ is a scalar product.
As $0(0,\vec{u}) = 0 \, \forall \vec{u}$ and $\Psi (0,\vec{u}) = 0 \, \forall
\vec{u}$, we get by \Theorref{theorem_2} that $\Psi = 0$. 
\end{proof}

We demand that the quantum states are elements of a Hilbert space. So we
must complete the pre-Hilbert space $(\hH, <.|.>_{\hH})$.

%
% ------------------------------ Definition -------------------------------------
%

\begin{definition}
Let
\begin{eqnarray}
\fl \cH & = & \Big\{ F : \bR^4 \to \bar{\bC}^4 \Big| \, 
 F \circ \sigma_\lambda \in L_2 (\bR^3)^4 \, \forall \lambda \in \cP\; 
\mbox{and} \; 
\exists \mbox{sequence} \, \{\Psi_m\}_{m\in\bN}, \, \Psi_m \in \hH : \nonumber \\
\fl & & \; \forall \epsilon > 0 \;\; \exists N_\epsilon :
 \fnorm{(F - \Psi_m) \circ \sigma_\lambda}_\lambda < \epsilon \quad
 \forall m>N_\epsilon \;\; \forall \lambda \in \cP \Big\}
\end{eqnarray}
Let $F \in \cH$, we define $F = 0 \Leftrightarrow \fnorm{F \circ
  \sigma_\lambda}_\lambda = 0 \, \forall \lambda \in \cP$.
\\
A scalar product $<.|.>_{\cH} : \cH \times \cH \to \bC$ is defined by:
\begin{eqnarray*}
<F_1 | F_2>_{\cH} \, := \, <F_1 \circ \sigma_\lambda | F_2 \circ \sigma_\lambda>_\lambda
 \quad , \quad \forall F_1, F_2 \in \cH
\end{eqnarray*}
with $\lambda \in \cP$ arbitrary.
\end{definition}

The following theorem proves that $(\cH, <.|.>_{\cH})$ is really a Hilbert space and
a completion of ${(\hH,<.|.>_{\hH})}$.

%
% ------------------------------ Theorem 4 --------------------------------------
%

\begin{theorem}
The above scalar product is well defined (independent of the parameter
$\lambda$). $(\cH,<.|.>_{\cH})$ is a Hilbert space and $\hH$ is a dense subspace of
it.
\end{theorem}

%
% ------------------------------ Proof of Theorem 4 -----------------------------
%

\begin{proof}
(i) We first prove that $\cH$ is a vector-space. The only thing which is
(perhaps) not trivial is the existence of a sequence in the above sense. 
Let $F_1,F_2 \in \cH$, $a,b \in \bC$, then it exits sequences $\Psi_{1,m}, \Psi_{2,m}$ in the
above sense. Now we get
\begin{eqnarray*}
& & \fnorm{((aF_1 + bF_2) - (a\Psi_{1,m}+b\Psi_{2,m})) \circ
  \sigma_\lambda}_\lambda \\
& = & \fnorm{a(F_1 - \Psi_{1,m}) \circ \sigma_\lambda + b (F_2 - \Psi_{2,m})
  \circ \sigma_\lambda}_\lambda \\
& \le & \fabs{a} \fnorm{(F_1 - \Psi_{1,m}) \circ \sigma_\lambda}_\lambda + \fabs{b} \fnorm{(F_2 -
  \Psi_{2,m}) \circ \sigma_\lambda}_\lambda \\
& \stackrel{m \to \infty}{\longrightarrow} & 0
\end{eqnarray*}
uniformly for all $\lambda \in \cP$.
\\
(ii) We now prove that $<F_1 \circ \sigma_\lambda | F_2 \circ
\sigma_\lambda>_\lambda$ is independent of $\lambda \in \cP$ for all $F_1,F_2
\in \cH$.
Let $F_1,F_2 \in \cH$, then it exists sequences $\Psi_{1,m}, \Psi_{2,m}$ in the
above sense. Now we get with $\lambda,\bar{\lambda} \in \cP$
\begin{eqnarray*}
& & <F_1 \circ \sigma_\lambda | F_2 \circ \sigma_\lambda>_\lambda \\
& = & <{\lim-\lambda}_{m\to\infty} (\Psi_{1,m} \circ \sigma_\lambda) | {\lim-\lambda}_{m \to \infty}
(\Psi_{2,m} \circ \sigma_\lambda)>_\lambda \\
& = & \lim_{m \to \infty} < \Psi_{1,m} \circ \sigma_\lambda | \Psi_{2,m} \circ
\sigma_\lambda >_\lambda \\
& \stackrel{\Theorref{theorem_1}}{=} & \lim_{m \to \infty} < \Psi_{1,m}
\circ \sigma_{\bar{\lambda}} | \Psi_{2,m} \circ \sigma_{\bar{\lambda}} >_{\bar{\lambda}} \\
& = & <{\lim-\bar{\lambda}}_{m\to\infty} (\Psi_{1,m} \circ \sigma_{\bar{\lambda}})
| {\lim-\bar{\lambda}}_{m \to \infty} (\Psi_{2,m} \circ
\sigma_{\bar{\lambda}})>_{\bar{\lambda}} \\
& = & <F_1 \circ \sigma_{\bar{\lambda}} | F_2 \circ
\sigma_{\bar{\lambda}}>_{\bar{\lambda}}
\end{eqnarray*}
$\lim-\lambda$ and $\lim-\bar{\lambda}$ means the limit concerning the norms
$\fnorm{.}_\lambda$ and $\fnorm{.}_{\bar{\lambda}}$ in the Hilbert
spaces $(L_2 (\bR^3)^4, <.|.>_\lambda)$ and $(L_2 (\bR^3)^4, <.|.>_{\bar{\lambda}})$ .
It is now clear that $<.|.>_{\cH}$ is a sesquilinear form with
$<F|F>_{\cH} \ge 0 \; \forall F \in \cH$.
\\
(iii) Let $F \in \cH$ with $<F|F>_{\cH} = 0$, then it follows
\begin{eqnarray*}
\begin{array}{cll}
& <F|F>_{\cH} = 0 \\
\Rightarrow & <F \circ \sigma_\lambda | F \circ \sigma_\lambda>_\lambda = 0
& \forall \lambda \in \cP \\
\Rightarrow & \fnorm{F \circ \sigma_\lambda}_\lambda = 0 
& \forall \lambda \in \cP \\
 \big(\Leftrightarrow & (F \circ \sigma_\lambda) (\vec{u}) = 0 & \mbox{for almost
  all} \, \vec{u} \in \bR^3, \forall \lambda \in \cP\big) \\
\stackrel{def.}{\Leftrightarrow} & F = 0 &
\end{array}
\end{eqnarray*}
So $<.|.>_{\cH}$ is a scalar product.
\\
(iv) We now prove that $\cH$ is complete.
Let $\{F_m\}$ be a Cauchy-sequence in $\cH$, so $\fnorm{(F_m - F_n) \circ
 \sigma_\lambda}_\lambda \stackrel{m,n \to \infty}{\longrightarrow} 0$ uniformly
 $\forall \lambda \in \cP$.
So it exists a sub-sequence $\{F_{n_k}\}$ with
\begin{eqnarray*}
\fnorm{(F_{n_{k+1}} - F_{n_k}) \circ \sigma_\lambda}_\lambda \le 2^{-k} \quad
 \forall k \; \forall \lambda \in \cP
\end{eqnarray*}
By using Lebesgue's dominated convergence theorem with the sequence
$f_{\lambda,k} := F_{n_k} \circ \sigma_\lambda$, we can show that it exist $f_\lambda \in
L_2(\bR^3)^4$ with
\begin{eqnarray*}
\fnorm{F_{n_k} \circ \sigma_\lambda - f_\lambda}_\lambda
\stackrel{k\to\infty}{\longrightarrow} 0 \qquad \forall \lambda \in \cP
\end{eqnarray*}
and $\lim_{k\to\infty} F_{n_k} \circ \sigma_\lambda (\vec{u}) =
f_\lambda (\vec{u})$ almost everywhere and $\forall \lambda \in
\cP$. Note that the sub-sequence $\{n_k\}_{k\in\bN}$ is independent of
$\lambda$!
We also get $\fnorm{F_n \circ \sigma_\lambda - f_\lambda}_\lambda
\stackrel{n\to\infty}{\longrightarrow} 0 \qquad \forall \lambda \in \cP$.
Because $\fnorm{(F_m - F_n) \circ
 \sigma_\lambda}_\lambda \stackrel{m,n \to \infty}{\longrightarrow} 0$ uniformly
 $\forall \lambda \in \cP$, we get by taking ${\lim-\lambda}_{m \to
 \infty}$ that $\fnorm{F_n \circ \sigma_\lambda - f_\lambda}_\lambda
\stackrel{n\to\infty}{\longrightarrow} 0$ uniformly $\forall \lambda
 \in \cP$.
Now we set
\begin{eqnarray*}
F (x) = \left\{ \begin{array}{cc} f_\lambda (\vec{u}) & \mbox{if} \;
\sigma_\lambda(\vec{u}) = x \; \mbox{and} \; \lim_{k\to\infty} F_{n_k} \circ
\sigma_\lambda(\vec{u}) = f_\lambda (\vec{u}) \\
0 & \mbox{otherwise} \end{array} \right.
\end{eqnarray*}
This function is well defined, because the sub-sequence $F_{n_k}$ is independent
of $\lambda$!
It is also trivial that $F \circ \sigma_\lambda = f_\lambda$ almost everywhere.
\\
We now prove that $F \in \cH$. The only thing which is left to prove is the
existence of a sequence $\Psi_m \in \hH$.
Because $F_m \in \cH$, it exists sequences $\{\Phi_{m,v}\}$ with
$\fnorm{(F_m - \Phi_{m,v}) \circ \sigma_\lambda}_\lambda
\stackrel{v \to \infty}{\longrightarrow} 0$ uniformly
 $\forall \lambda \in \cP$.
So it exists $\Psi_m \in \hH$ with
$\fnorm{(F_m - \Psi_m) \circ \sigma_\lambda}_\lambda < \frac{1}{m} \qquad \forall \lambda \in \cP$.
Now we get
\begin{eqnarray*}
& & \fnorm{(F - \Psi_m) \circ \sigma_\lambda}_\lambda \\
& \le & \fnorm{(F - F_m) \circ \sigma_\lambda}_\lambda + \fnorm{(F_m - \Psi_m)
  \circ \sigma_\lambda}_\lambda \\
& \le & \underbrace{\fnorm{f_\lambda - F_m \circ \sigma_\lambda}_\lambda}_{\stackrel{m\to\infty}{\longrightarrow}
  0 \; \mbox {uniformly} \; \forall \lambda \in \cP} + \frac{1}{m}
\stackrel{m\to\infty}{\longrightarrow} 0
\end{eqnarray*}
uniformly(!) for all $\lambda \in \cP$.
So it results that $F \in \cH$.
\\
The last step to prove is: $F_m
\stackrel{m\to\infty}{\longrightarrow} 0$ concerning the norm in $\cH$. We get
for all $\lambda \in \cP$:
\begin{eqnarray*}
\fnorm{(F - F_m) \circ \sigma_\lambda}_\lambda = \fnorm{f_\lambda -
F_m \circ \sigma_\lambda}_\lambda
\stackrel{m\to\infty}{\longrightarrow} 0
\end{eqnarray*}
(v) It is trivial that $\hH \subset \cH$ and that $\hH$ is dense in $\cH$.
\end{proof}

We are now in position to postulate that the (pure) states of the
quantum part of the total system are the elements of the Hilbert space
$(\cH, <.|.>_{\cH})$.

Let $\lambda \in \cP$ and we define the function $U_\lambda : \cH \to
\cR_\lambda$ by
\begin{eqnarray}
U_\lambda : \cH \ni F (x) \longmapsto F \circ \sigma_\lambda
(\vec{u}) \in \cR_\lambda
\label{sec2_U}
\end{eqnarray}
with $\cR_\lambda \subset L_2(\bR^3)^4$ denoting the range of
$U_\lambda$.
A quantum state is uniquely given by its values on a hyperplane $\sigma_\lambda$. This means
that the function $U_\lambda$ is injective for all $\lambda \in
\cP$. The following theorem proves this property.

%
% ------------------------------ Theorem 5 --------------------------------------
%

\begin{theorem}
Let $F_1, F_2 \in \cH$ such that it exists $\mu \in \cP$ with
$F_1 \circ \sigma_\mu = F_2 \circ \sigma_\mu$
then it follows
$F_1 = F_2$.
\end{theorem}

%
% ------------------------------ Proof of Theorem 5 -----------------------------
%

\begin{proof}
Let $F = F_1 - F_2$, we get $F \circ \sigma_\mu = 0$. $F \in \cH$, so it exists sequence
$\Psi_n \in \hH$ with
$\fnorm{(F-\Psi_n) \circ \sigma_\lambda}_\lambda
\stackrel{n\to\infty}{\longrightarrow} 0$
uniformly for all $\lambda \in \cP$.
We get
\begin{eqnarray*}
\fl 0 \stackrel{n \to \infty}{\longleftarrow} \fnorm{(F - \Psi_n) \circ
  \sigma_\mu}_\mu = \fnorm{\Psi_n \circ \sigma_\mu}_\mu 
\stackrel{\Theorref{theorem_1}}{=} \fnorm{\Psi_n \circ \sigma_\lambda}_\lambda
  \; , \quad \forall \lambda \in \cP
\end{eqnarray*}
Because
\begin{eqnarray*}
\fnorm{F \circ \sigma_\lambda}_\lambda \le
\fnorm{(F - \Psi_n) \circ \sigma_\lambda}_\lambda + \fnorm{\Psi_n \circ
    \sigma_\lambda}_\lambda \stackrel{n\to\infty}{\longrightarrow} 0 
\quad \forall \lambda \in \cP
\end{eqnarray*}
we get $\fnorm{F \circ \sigma_\lambda}_\lambda = 0 \, \forall
\lambda \in \cP \Leftrightarrow F = 0$.\\
It follows that $0 = F = F_1 - F_2 \, \Rightarrow \, F_1 = F_2$.
\end{proof}

The function $U_\lambda$ is invertible, let $U^{-1}_\lambda :
\cR_\lambda \to \cH$ be the inverse function. The following theorem
proves some properties of $U_\lambda$ and $U^{-1}_\lambda$
respectively.

%
% ------------------------------ Theorem 6 --------------------------------------
%

\begin{theorem}
Let $\lambda \in \cP$ and the functions $U_\lambda : \cH \to
\cR_\lambda$ and $U^{-1}_\lambda : \cR_\lambda \to \cH$ are defined as
above.
\begin{itemize}
\item[(i)] Let $F \in \cH$ and $f \in \cR_\lambda$, then
$<U^{-1}_\lambda f | F>_{\cH} \, = \, <f|U_\lambda F>_\lambda$
and especially
$\fnorm{U_\lambda F}_\lambda = \fnorm{F}_{\cH}$ and
$\fnorm{U^{-1}_\lambda f}_{\cH} = \fnorm{f}_\lambda$

\item[(ii)] We use the Hilbert space $(L_2(\bR^3)^4,
<.|.>_\lambda)$. Then the set $\cR_\lambda \subset L_2 (\bR^3)^4$ is closed.
\end{itemize}
\end{theorem}

%
% ------------------------------ Proof of Theorem 6 -----------------------------
%

\begin{proof}
(i) 
$<U^{-1}_\lambda f | F>_{\cH} 
= <\underbrace{(U^{-1}_\lambda f) \circ \sigma_\lambda}_{f} |
     \underbrace{F \circ \sigma_\lambda}_{U_\lambda F}>_\lambda 
= < f | U_\lambda F>_\lambda$
\\
(ii)
Let $f_n \in \cR_\lambda$ for all $n\in\bN$ and $\lim-\lambda_{n\to\infty}
f_n = f \in L_2(\bR^3)^4$. We want to prove that $f \in
\cR_\lambda$. We set $F_n := U^{-1}_\lambda f_n$. 
$\{F_n\}_{n\in\bN}$ is also a Cauchy-sequence. Because $\cH$ is
complete, it exists $F \in \cH$ with $\lim_{n\to\infty} F_n =
F$. Moreover we get
\begin{eqnarray*}
f = \lim-\lambda_{n\to\infty} f_n = \lim-\lambda_{n\to\infty} U_\lambda F_n 
= U_\lambda \lim_{n\to\infty} F_n = U_\lambda F
\end{eqnarray*}
(because $U_\lambda$ is bounded/continuous).
So $\cR_\lambda$ is closed.
\end{proof} 

It follows that $(\cR_\lambda, <.|.>_\lambda)$ is a Hilbert space and
$U_\lambda : (\cH,<.|.>_{\cH}) \to (\cR_\lambda, <.|.>_\lambda)$ is an
unitary operator for all $\lambda \in \cP$.
Because $(L_2 (\bR^3)^4, <.|.>_\lambda)$ is a separable Hilbert space,
$(\cR_\lambda, <.|.>_\lambda)$ is a separable ``sub''-Hilbert
space. Therefore, $(\cH, <.|.>_{\cH})$ must be a separable Hilbert space.

% ===============================================================================
% Section 3
% Change of the reference frame and charge conjugation
% ===============================================================================

\section{Change of the reference frame and Charge conjugation}

Our aim is now to define how the quantum state changes, if we
change the reference frame $K \to \widetilde{K}$ with $\widetilde{x} = \Lambda x + a$.
The classical state does not change in this case.

We look only at Poincar\'e-transformations $(\Lambda, a)$ which do not mirror
the space and do not invert the direction of time, i.e., $\Lambda \in
L_+^\uparrow$. Let $S (\Lambda)$ be a non-singular $4 \times
4$-matrix with $S^{-1}(\Lambda) \gamma^\mu S(\Lambda) =
\Lambda^\mu_{\;\;\nu}~\gamma^\nu$, $S^{-1}(\Lambda) = S(\Lambda^{-1})$
and $S^{-1}(\Lambda) = \gamma^0 S^+(\Lambda) \gamma^0$.

Let us first present a lemma which will be needed in the proofs of the main
theorems.

%
% ------------------------------ Lemma  -----------------------------------------
%

\begin{lemma}
Let $f,g : \bR^4 \to \bar{\bC}^4$, $f \circ \sigma_\lambda, g \circ
\sigma_\lambda \in L_2(\bR^3)^4$ for all $\lambda \in \cP$, $\Lambda
\in L_+^\uparrow$, $a \in \bR^4$, we set
\begin{eqnarray*}
\widetilde{f} (\widetilde{x}) = S(\Lambda) f (\Lambda^{-1} (\widetilde{x} - a)) \\
\widetilde{g} (\widetilde{x}) = S(\Lambda) g (\Lambda^{-1} (\widetilde{x} - a))
\end{eqnarray*}
Let $\lambda \in \cP$ arbitrary, then it exists $\mu(\lambda) \in \cP$ with
\begin{eqnarray*}
<\widetilde{f} \circ \sigma_\lambda | \widetilde{g} \circ \sigma_\lambda>_\lambda
\, = \, <f \circ \sigma_{\mu(\lambda)} | g \circ \sigma_{\mu(\lambda)}
>_{\mu(\lambda)}
\end{eqnarray*}
\end{lemma}

%
% ------------------------------ Proof of Lemma ---------------------------------
%

\begin{proof}
Each arbitrary Lorentz-transformation $\Lambda \in L^\uparrow_+$ can
be expressed as a product of pure translations, pure rotations
and Lorentz-boosts in the $x^1$-direction. So it is enough to prove
the lemma for pure translations, pure rotations and
Lorentz-boosts in the $x^1$-direction separately. This can be done by
straightforward calculations.
\end{proof}

The electromagnetic potential in the reference frame $\tilde{K}$ is given
by 
\begin{eqnarray*}
\widetilde{A}_\mu (\widetilde{x}) = (\Lambda^{-1})^\nu_{\;\;\mu}
A_\nu (\Lambda^{-1}(\tilde{x}-a))
\end{eqnarray*}
So we define
\begin{eqnarray}
\fl \widetilde{\hH} = \Big\{ & 
\widetilde{\Psi} : \bR \times \bR^3 \to \bC^4 \Big|
\widetilde{\Psi}  \; \mbox{cont. diff.}, 
\left(\rmi \gamma^\mu \partial_\mu - \frac{e}{c\hbar} \gamma^\mu \widetilde{A}_\mu -
    \frac{mc}{\hbar} \right) \widetilde{\Psi} (\widetilde{x}) = 0, \nonumber\\
\fl & \fnorm{\widetilde{\Psi} \circ \sigma_\lambda}_\lambda < \infty \, ,
\; \lim_{\fabs{\vec{u}} \to \infty} \fabs{\vec{u}}^3 
\fabsq{\widetilde{\Psi} \circ \sigma_\lambda (\vec{u})} = 0 \quad \forall
\lambda \in \cP \Big\}
\label{sec3_def_tilde_hH}
\end{eqnarray}

A scalar product $<.|.>_{\widetilde{\hH}}$ between two elements of
$\widetilde{\hH}$ and a completion $(\widetilde{\cH},
<.|.>_{\widetilde{\cH}})$ can be constructed in the
same way as in the previous section.

Let the quantum state in the reference frame $K$ be $\Psi \in
\hH$. Then the quantum state in the reference frame $\widetilde{K}$ is
defined to be
\begin{eqnarray*}
\widetilde{\Psi} (\widetilde{x}) = S(\Lambda) \Psi (\Lambda^{-1} (\widetilde{x} - a))
\end{eqnarray*}
We get the following theorem:

%
% ------------------------------ Theorem 7 --------------------------------------
%

\begin{theorem}
Let $\Psi \in \hH$, then
$\widetilde{\Psi} (\widetilde{x}) = 
S(\Lambda) \Psi (\Lambda^{-1} (\widetilde{x} - a)) \in \widetilde{\hH}$.
\label{theorem_cov1}
\end{theorem}

%
% ------------------------------ Proof of Theorem 7 -----------------------------
%

\begin{proof}
It is clear that $\widetilde{\Psi}$ is continuous differentiable and
that $\widetilde{\Psi}$ is a
solution of the Dirac-equation with external field $\widetilde{A}_\mu$ (see for
example \cite{bjorken.book}). The third condition in
\eqref{sec3_def_tilde_hH} is clear because of the lemma.
The last condition can be proved by simple calculations.
Again, it is enough to do this only for pure
translations, pure rotations and Lorentz-boosts in the $x^1$-direction
separately.
\end{proof}

Now, we look at the general case $F \in \cH$. Let us define an
operator $W_{(\Lambda,a)} : \cH \to \widetilde{\cH}$:
\begin{eqnarray*}
(W_{(\Lambda,a)} F) (\widetilde{x}) :=  S(\Lambda) F (\Lambda^{-1} (\widetilde{x} -
a)) \quad \forall F \in \cH
\end{eqnarray*}
This operator is well defined. Using these transformation rules, the
scalar product is covariant. Its value is
equal in all reference frames. Or in other words: the operator $W_{(\Lambda,a)}$
is unitary. All these will be proven by the next theorem.

%
% ------------------------------ Theorem 8 --------------------------------------
%

\begin{theorem}
\begin{itemize}
\item[(i)] Let $F \in \cH$, then
$\widetilde{F} (\widetilde{x}) = 
(W_{(\Lambda,a)} F) (\widetilde{x}) \in \widetilde{\cH}$.

\item[(ii)] Let $F_1, F_2 \in \cH$, define $\widetilde{F_1} = W_{(\Lambda,a)} F_1$,
$\widetilde{F_2} = W_{(\Lambda,a)} F_2$
then
\begin{eqnarray*}
<\widetilde{F_1}|\widetilde{F_2}>_{\widetilde{\cH}} 
\; = \; <W_{(\Lambda,a)} F_1 | W_{(\Lambda,a)} F_2>_{\widetilde{\cH}}
\; = \; <F_1|F_2>_{\cH}
\end{eqnarray*}
\end{itemize}
\label{theorem_cov2}
\end{theorem}

%
% ------------------------------ Proof of Theorem 8 -----------------------------
%

\begin{proof}
(i) Because of the lemma, we get
\begin{eqnarray*}
\fnorm{\widetilde{F} \circ \sigma_\lambda}_\lambda =
\fnorm{F \circ \sigma_{\mu(\lambda)}}_{\mu(\lambda)} < \infty
\end{eqnarray*}
for all $\lambda \in \cP$.
\\
The existence of the sequence $\widetilde{\Psi}_n \in \widetilde{\hH}$
is only left to prove.
Since $F \in \cH$, it exists a sequence
$\Psi_n \in \hH$ with $\fnorm{(F-\Psi_n) \circ \sigma_\lambda}_\lambda
\stackrel{n\to\infty}{\longrightarrow} 0$ uniformly for all $\lambda \in \cP$.
Let $\widetilde{\Psi}_n (\widetilde{x}) = S(\Lambda) \Psi_n (\Lambda^{-1} (\widetilde{x} -
a))$. $\widetilde{\Psi}_n \in \widetilde{\hH}$ because of \Theorref{theorem_cov1}.
Since $\fnorm{(F - \Psi_n) \circ \sigma_\lambda}_\lambda \stackrel {n \to
\infty}{\longrightarrow} 0$ uniformly for all $\lambda \in \cP$, we
also get
\begin{eqnarray*}
\fnorm{(\widetilde{F} - \widetilde{\Psi}_n) \circ \sigma_\lambda}_\lambda 
\stackrel{\mbox{see lemma}}{=} \fnorm{(F - \Psi_n) \circ
  \sigma_{\mu(\lambda)}}_{\mu(\lambda)}
\stackrel {n \to \infty}{\longrightarrow} 0
\end{eqnarray*}
uniformly for all $\lambda \in \cP$.
We have indeed $\widetilde{F} \in \widetilde{\cH}$.
\\
(ii)
\begin{eqnarray*}
<\widetilde{F_1}|\widetilde{F_2}>_{\widetilde{\cH}} & = &
<\widetilde{F_1} \circ \sigma_\lambda |\widetilde{F_2} \circ \sigma_\lambda>_\lambda \;
, \quad \mbox{arbitrary} \, \lambda \in \cP \\
& \stackrel{\mbox{see lemma}}{=} & <F_1 \circ \sigma_{\mu(\lambda)} |F_2 \circ
\sigma_{\mu(\lambda)}>_{\mu(\lambda)} \\
& = & <F_1 | F_2>_{\cH}
\end{eqnarray*}
\end{proof}

Now, we examine the situation if we charge conjugate the system $K \to
K^C$. We define
\begin{eqnarray}
\fl \hH^C = \Big\{ & 
\Psi^C : \bR \times \bR^3 \to \bC^4 \Big| 
\Psi^C \; \mbox{cont. diff.}, 
\left(\rmi \gamma^\mu \partial_\mu + \frac{e}{c\hbar} \gamma^\mu A_\mu -
    \frac{mc}{\hbar} \right) \Psi^C (x) = 0, \nonumber \\
\fl & \fnorm{\Psi^C \circ \sigma_\lambda}_\lambda < \infty \, ,
\; \lim_{\fabs{\vec{u}} \to \infty} \fabs{\vec{u}}^3 
\fabsq{\Psi^C \circ \sigma_\lambda (\vec{u})} = 0 \quad \forall
\lambda \in \cP \Big\}
\end{eqnarray}

Again a scalar product $<.|.>_{\hH^C}$ between two elements of
$\hH^C$ and a completion ${(\cH^C,<.|.>_{\cH^C})}$ can be constructed
in the same way as in the previous section.

It is well known that in any representation of the $\gamma$-matrices there must
exist a matrix $C$ which satisfies
\begin{eqnarray}
C {\gamma^\mu}^T C^{-1} = - \gamma^\mu
\label{sec3_prop_C}
\end{eqnarray}
(see e.g. \cite{bjorken.book}).
In addition, we want to use only representations of the $\gamma$-matrices for which
there exists an unitary matrix $C$ satisfying \eqref{sec3_prop_C}. (This
is true e.g. in the Dirac-representation with $C = \rmi \gamma^2 \gamma^0$.)

The following theorem expresses the relation between the
spaces $\cH$ and $\cH^C$. It can be proved by straightforward
calculations.

%
% ------------------------------ Theorem 9 --------------------------------------
%

\begin{theorem}
\begin{itemize}
\item [(i)]   Let $\Psi \in \hH$, then $\Psi^C = C {\gamma^0}^T \Psi^* \in \hH^C$.
\item [(ii)]  Let $F \in \cH$, then $F^C = C {\gamma^0}^T F^* \in \cH^C$.
\item [(iii)] Let $F_A, F_B \in \cH$, let $F_A^C = C {\gamma^0}^T F_A^*$,
$F_B^C = C {\gamma^0}^T F_B^*$, then
\begin{eqnarray*}
<F_A^C | F_B^C>_{\cH^C} \, = \, <F_B|F_A>_{\cH}
\end{eqnarray*}
\end{itemize}
\end{theorem}

% ===============================================================================
% Section 4
% Events Generating Algorithm (ideal, infinitesimal short measurements)
% ===============================================================================

\section{Events Generating Algorithm\\(ideal, infinitesimal short
measurements)}

In the previous sections, we have precisely defined the state of the
total system and examined some of its properties.
We are now in position to present the proper-time evolution of the
system state. More precisely, we will postulate algorithms which
generate events, i.e. irreversible changes of the classical state. 
Because we know that the set of quantum states is
indeed a Hilbert space, we can use the well-known formulation in
Hilbert space framework. 
In this section, we formulate an algorithm to describe ideal
measurements of infinitesimal short duration. In principle, we rewrite
the standard reduction 
postulate of the non-relativistic quantum mechanics by replacing $t$ with $\tau$
and using our Hilbert space of ``solutions.'' Doing this, we get a
covariant algorithm playing the role of the standard reduction postulate in
non-relativistic quantum mechanics.

We name the reference frame $K$. Let the particle be prepared at
proper time $\tau_0$ in a space-time point $z_0$.

There should be $n$ measurements, which happen at proper times $\tau_i$ in
space-time points $z_i$, $i=1..n$. The $i$th measurement is represented by an
observable $M_i$ with
\begin{eqnarray*}
  M_i = \sum_j \lambda_{i,j} \ket{\Phi_{i,j}}\bra{\Phi_{i,j}}
\end{eqnarray*}
$\Phi_{i,j} \in \cH$, $1 = \sum_j \ket{\Phi_{i,j}}\bra{\Phi_{i,j}}$,
$<\Phi_{i,j}|\Phi_{i,k}>_{\cH} = \delta_{j,k}$ and $\lambda_{i,j} \in \bR$.

We assume that $\tau_0 < \tau_1 < ... < \tau_n$.
We want to preserve a weak kind of order, so we demand the following:
no event (e.g. preparation, measurement or detection) can take place
in the backward light-cone of the previous event:
\begin{eqnarray*}
\fl (\fnormq{z_{j+1} - z_j} \ge 0 \; \mbox{and} \; z_j^0 <  z_{j+1}^0)
  \; \mbox{or} \; (\fnormq{z_{j+1} - z_j} < 0) \quad \forall j=0,1,..n-1
\end{eqnarray*}
with $\fnormq{x} = \fnormq{(x^0,\vec{x})} = (x^0)^2 - \fabsq{\vec{x}}$ being the
Minkowski-distance.
Or in other words: let two successive events happen in space-time points $z_j = (z_j^0,
\vec{z}_j)$ and $z_{j+1}= (z_{j+1}^0, \vec{z}_{j+1})$, then it must
exist a Poincar\'e-transformation $(\Lambda, a)$ ($\Lambda
\in L_+^\uparrow$) such that
\begin{eqnarray*}
\Lambda^0_{\;\mu} z_j^\mu + a^0 = \tilde{z}_j^0 \stackrel{!}{<} \tilde{z}_{j+1}^0 
= \Lambda^0_{\;\mu} z_{j+1}^\mu + a^0
\end{eqnarray*}
It must exist a reference frame in which the time of the
first event $\tilde{z}_j^0$ is earlier than the time of 
the second event $\tilde{z}_{j+1}^0$.

Now we start with the formulation of a relativistic reduction-postulate
for ideal measurements. Let $(\omega_\tau, \Psi_\tau)$ be the state of
the total system.
\begin{itemize}
\item[(i)] The particle is prepared at proper time $\tau_0$ in space-time point
  $z_0$. The quantum state is given by $\Psi_{\tau_0}$ 
  with $\fnormq{\Psi_{\tau_0}}_{\cH} = 1$ and the classical state is
  $\omega_{\tau_0} = 0$. Let $i=1$.

\item[(ii)] The quantum and classical state change only in case of measurement. They have
no $\tau$-dependence if there is no measurement:
\begin{eqnarray*}
  (\omega_\tau, \Psi_\tau ) =  (\omega_{\tau_{i-1}}, \Psi_{\tau_{i-1}})
\end{eqnarray*}
for $\tau_{i-1} \le \tau \le \tau_{i}$. 

\item[(iii)] The $i$th measurement takes place at proper time $\tau_i$ in
  space-time point $z_i$. We choose
  the measurement result $\lambda_{i,j}$ with probability
\begin{eqnarray*}
  p(\lambda_{i,j}) = \fabsq{<\Phi_{i,j}|\Psi_{\tau_i}>_{\cH}}
\end{eqnarray*}
If $\lambda_{i,j}$ is the received measurement result, the state of the total system
changes in the following way:
\begin{eqnarray*}
  (\omega_{\tau_i},\Psi_{\tau_i})  \longrightarrow (j,\Phi_{i,j})
\end{eqnarray*}

\item[(iv)] Let $i \to i+1$ and go to step (ii).
\end{itemize}

We want to examine how this algorithm looks like in another reference frame.
Let $\widetilde{K}$ be a reference frames which is connected to $K$ by a
Poincar\'e-transformation $(\Lambda,a)$ with $\Lambda \in L_+^\uparrow$.

In $\widetilde{K}$, the situation is described in this way: the particle
is prepared at $\tau_0$ in $\widetilde{z}_0 = \Lambda z_0 + a$ with
initial quantum state
\begin{eqnarray*}
\widetilde{\Psi}_{\tau_0} (\widetilde{x}) & = & (W_{(\Lambda,a)} \Psi_{\tau_0}) (\widetilde{x}) =
S(\Lambda) \Psi_{\tau_0} (\Lambda^{-1}(\widetilde{x}-a))
\end{eqnarray*}
with $\fnormq{\widetilde{\Psi}_{\tau_0}}_{\widetilde{\cH}} = 1$ (the operator
$W_{(\Lambda,a)}$ is unitary).
The measurement $i$ at proper time $\tau_i$ happens in
$\widetilde{z}_i = \Lambda z_i + a$ and is
represented by 
\begin{eqnarray*}
\widetilde{M}_i = W_{(\Lambda,a)} M_i  W_{(\Lambda,a)}^+ 
= \sum_j \lambda_{i,j} \ket{\widetilde{\Phi}_{i,j}}\bra{\widetilde{\Phi}_{i,j}}
\end{eqnarray*}
with $\widetilde{\Phi}_{i,j} = W_{(\Lambda,a)} \Phi_{i,j}$.
It is true that $1 = \sum_j \ket{\widetilde{\Phi}_{i,j}}\bra{\widetilde{\Phi}_{i,j}}$ and
$<\widetilde{\Phi}_{i,j}|\widetilde{\Phi}_{i,k}>_{\widetilde{\cH}} = \delta_{j,k}$.

If we apply the algorithm in $\widetilde{K}$ and if we choose the same random
numbers, then we get the same measurement results than those we get if we apply
the algorithm in $K$, because
\begin{eqnarray*}
\fl \tilde{p}(\lambda_{i,j})
= \fabsq{<W_{(\Lambda,a)} \Phi_{i,j}|W_{(\Lambda,a)} \Psi_{\tau_i}>_{\widetilde{\cH}}} 
= \fabsq{<\Phi_{i,j}|\Psi_{\tau_i}>_{\cH}}
= p(\lambda_{i,j}) 
\end{eqnarray*}

The system state $(\omega_\tau, \Psi_\tau)$ in the reference frame $K$ and
the system state $(\widetilde{\omega}_\tau, \widetilde{\Psi}_\tau)$ in the reference frame
$\widetilde{K}$ are always connected in the following way:
\begin{eqnarray*}
(\widetilde{\omega}_\tau, \widetilde{\Psi}_\tau) = 
(\omega_\tau, W_{(\Lambda,a)} \Psi_\tau)
\end{eqnarray*}

The above algorithm describing ideal, infinitesimal short measurements
is covariant.

Now we consider the charge conjugated system $K^C$. We set
\begin{eqnarray*}
\Psi_{\tau_0}^C    & := & C{\gamma^0}^T \Psi_{\tau_0}^* \\
\Phi_{i,j}^C       & := & C{\gamma^0}^T \Phi_{i,j}^*
\end{eqnarray*}
The charge conjugated observables are defined by
\begin{eqnarray*}
M_i^C  = C {\gamma^0}^T M_i^* {\gamma^0}^T C^+ 
=  \sum_j \lambda_{i,j} \ket{\Phi_{i,j}^C}\bra{\Phi_{i,j}^C}
\end{eqnarray*}
The complex conjugated operator $M_i^*$ is defined by $M_i^* \Psi := (M_i
\Psi^*)^*$.

If we execute the
algorithm in a charge conjugated system $K^C$ or if we execute the algorithm in the
normal system $K$, both will result the same events (if we choose the same random
numbers), because
\begin{eqnarray*}
p^C (\lambda_{i,j}) = \fabsq{<\Phi_{i,j}^C | \Psi_{\tau_i}^C>_{{\cH}^C}} = 
\fabsq{<\Psi_{\tau_i} | \Phi_{i,j}>_{\cH}} = p (\lambda_{i,j})
\end{eqnarray*}

The system state $(\omega_\tau, \Psi_\tau)$ in $K$ and
the system state $(\omega^C_\tau, \Psi_\tau^C)$ in $K^C$ are always connected by
\begin{eqnarray*}
(\omega^C_\tau, \Psi_\tau^C ) = (\omega_\tau, C {\gamma^0}^T \Psi_\tau^*) 
\end{eqnarray*}

We demand that the algorithm applied in the ``charge conjugated world'' or applied in the
``normal world'' describes the same physical situation. 

We end this section with the derivation of an important relationship between the
standard reduction postulate used with the Dirac-equation and the above algorithm:
the standard reduction 
postulate formulated in a (preferred) fixed reference frame can be rewritten as
a special case of the above algorithm. Especially, the standard reduction 
postulate used in a fixed reference frame gives the same probabilities than
the above (covariant) algorithm.

We choose the (preferred) fixed reference frame. We assume that the electromagnetic
potential $A_\mu$ is time-independent in this frame. Now, we define
\begin{eqnarray*}
H_D = -\rmi c \hbar \gamma^0 \gamma^k \frac{\partial}{\partial x^k} + e
\gamma^0 \gamma^\mu A_\mu + \gamma^0 mc^2
\end{eqnarray*}

Let $U_t \equiv  U_{((ct,\vec{0}),\vec{0},\vec{0})}$ (see
\eqref{sec2_U}), so that $(U_t \Psi)(\vec{u}) = \Psi (ct,\vec{u})$.
We are now in position to prove our claim.

Let the wave function be $\psi_0$ at
time $t=0$ with $\fnorm{\psi_0}_{L_2(\bR^3)^4} = 1$. We assume measurements happening at
times $t_1,..,t_n$ with $0 < t_1 < t_2 < ... < t_n$. The measurement
$i$ is represented by an observable $m_i$ with
\begin{eqnarray*}
  m_i = \sum_j \lambda_{i,j} |\phi_{i,j}><\phi_{i,j}|
\end{eqnarray*}
and $1 = \sum_j |\phi_{i,j}><\phi_{i,j}|$,
$<\phi_{i,j}|\phi_{i,k}>_{L_2(\bR^3)^4} = \delta_{j,k}$ and
$\lambda_{i,j} \in \bR$.

Next, we describe this situation in the framework of our formalism. Let
\begin{eqnarray*}
\Psi_0 & := & U^{-1}_0 \psi_0
\end{eqnarray*}
We get $\fnorm{\Psi_0}_{\cH} = 1$.
We define $n$ measurements happening at proper times $\tau_i := t_i$ at space-time
points $z_i = (ct_i, \vec{y}_i)$. $\vec{y}_i$ can be chosen arbitrary.
The measurements are represented by observables $M_i$ with
\begin{eqnarray*}
M_i := U^{-1}_{ct_i} m_i U_{ct_i}
= \sum_j \lambda_{i,j} |\Phi_{i,j}><\Phi_{i,j}|
\end{eqnarray*}
with $\Phi_{i,j} = U^{-1}_{ct_i} \phi_{i,j}$. Note that 
$1 = \sum_j |\Phi_{i,j}><\Phi_{i,j}|$ and $<\Phi_{i,j}|\Phi_{i,k}>_{\cH} = \delta_{j,k}$.

We execute the standard reduction postulate (SR) and the above
algorithm (AL):
\begin{itemize}
\item[(i)]
\begin{itemize}
\item[SR:] At time $t=0$ the wave function is $\psi_0$.

\item[AL:] At $\tau=0$ the state of the quantum part is $\Psi_0$ with
$\psi_0 = U_0 \Psi_0$.
\end{itemize}

\item[(ii)]
\begin{itemize} 
\item[SR:] Until $t=t_1$, the time evolution of the wave function is given by
\begin{eqnarray*}
\psi (t) = \fexp{-\frac{\rmi}{\hbar} t \, H_D} \psi_0
\end{eqnarray*}

\item[AL:] The state of the quantum part does not change until $\tau = \tau_1 =
t_1$:
\begin{eqnarray*}
\Psi_{\tau} = \Psi_0
\end{eqnarray*}
\end{itemize}
The following relationship between $\psi(t)$ and $\Psi_0$ is fulfilled
for $0 \le t \le t_1$:
\begin{eqnarray*}
\psi (t) = \fexp{-\frac{\rmi}{\hbar} t \, H_D} U_0 \Psi_0
= U_t \Psi_0 = U_t \Psi_{t}
\end{eqnarray*}

\item[(iii)]
\begin{itemize}
\item[SR:] At $t=t_1$, the first measurement happens. The probability for
  the result $\lambda_{1,j}$ is given by
\begin{eqnarray*}
p_{1,j} = \fabsq{<\phi_{1,j}| \psi(t_1)>_{L_2(\bR^3)^4}}
\end{eqnarray*}

\item[AL:] At $\tau = \tau_1 = t_1$, the first measurement happens. The probability for
the result $\lambda_{1,j}$ is given by
\begin{eqnarray*}
p_{1,j} & = & \fabsq{<\Phi_{1,j}| \Psi_0>_{\cH}} \\
& = & \fabsq{<U^{-1}_{t_1} \phi_{1,j}|
  U^{-1}_{t_1} \psi(t_1)>_{\cH}} \\
& = & \fabsq{<\phi_{1,j}|\psi (t_1)>_{L_2(\bR^3)^4}}
\end{eqnarray*}
\end{itemize}

\item[(iv)]
\begin{itemize}
\item[SR:] The result should be $\lambda_{1,j}$. Then, the following change
  of the wave function happens
\begin{eqnarray*}
\psi (t_1) \longrightarrow \phi_{1,j} = U_{t_1} \Phi_{1,j}
\end{eqnarray*}

\item[AL:] The result should be $\lambda_{1,j}$. Then, the following change
  of the wave function happens
\begin{eqnarray*}
\Psi_0 \longrightarrow \Phi_{1,j} = U^{-1}_{t_1} \phi_{1,j}
\end{eqnarray*}
\end{itemize}
\end{itemize}
The algorithm continues with the other measurements. 

We want to underline two
facts. First the probabilities resulting from the standard reduction postulate and
the probabilities resulting from the above algorithm are equal. Additionally, it is true
for all $t \ge 0$ that
\begin{eqnarray*}
\psi (t) = U_t \Psi_{t}
\end{eqnarray*}

% ===============================================================================
% Section 5
% Events Generating Algorithm (detections of the particle)
% ===============================================================================

\section{Events Generating Algorithm (detections of the particle)}

In this section, we formulate an algorithm for modelling continuous relativistic
measurements, indeed we 
will propose in the following an algorithm to simulate detections of the particle.
In principle, we will do this rewriting the algorithm of EEQT by
replacing $t$ with $\tau$ and using our Hilbert space of
``solutions.''

We label the reference frame $K$. The particle is prepared at proper
time $\tau_0$ in a point $x_0=(x_0^0,\vec{x_0})$.

We consider $n$ detectors with trajectories $z_j (\tau)$, $j=1..n$. The
trajectories start at proper time $\tau = \tau_0$ from the backward light-cone of the
space-time point of the 'preparation event':
\begin{eqnarray*}
\fnormq{x_0 - z_j (\tau_0)} = 0, \quad z_j^0(\tau_0) \le x_0^0
\end{eqnarray*}
 
We allow detections which happen in the past of the preparation time.
But we do not allow detections, if the detection space-time point is located in the
backward light-cone of the space-time point of the preparation event.

Each detector is characterized by operators $G_j (\tau)$. Let
$G_j^+ (\tau)$ be the adjoint operators. The total coupling between the quantum
and the classical system is given by $\Lambda(\tau) := \sum_j G_j^+ (\tau) G_j (\tau)$.

Let $(\omega_\tau, \Psi_\tau)$ be the state of
the total system. We define the following algorithm:
\begin{itemize}
\item[(i)] The particle is prepared in a space-time point $x_0$ at proper time
  $\tau=\tau_0$. The quantum state is $\Psi_{\tau_0}$ with
  $\fnormq{\Psi_{\tau_0}}_{\cH} = 1$ and
  the classical state is $\omega_{\tau_0} = 0$.

\item[(ii)] Choose a uniform random number $r \in [0,1]$.

\item[(iii)] Propagate the quantum state forward in proper time by solving
\begin{eqnarray}
  \frac{\partial}{\partial \tau} \Psi_\tau = -\frac{1}{2} \Lambda(\tau)
  \Psi_\tau
\label{sec5_dgl}
\end{eqnarray}
until $\tau = \tau_1$, where $\tau_1$ is defined by
\begin{eqnarray*}
  1 - \fnormq{\Psi_{\tau_1}}_{\cH} \, = \int_{\tau_0}^{\tau_1} \rmd\tau
  <\Psi_\tau|\Lambda\Psi_\tau>_{\cH} \, = r
\end{eqnarray*}
Let $\omega_\tau = \omega_{\tau_0}$ until $\tau=\tau_1$, a
detection happens at proper time $\tau = \tau_1$.

\item[(iv)] We choose the detector $k$ - which detects the particle - with
  probability
\begin{eqnarray*}
  p_k = \frac{1}{N} \fnormq{G_k(\tau_1) \Psi_{\tau_1}}_{\cH}
\end{eqnarray*}
with $N = \sum_j \fnormq{G_j (\tau_1)\Psi_{\tau_1}}_{\cH}$.

\item[(v)] Let $l$ be the detector which detects the particle. The
  detection happens at the point $z_l (\tau_1)$. The detection induces the
  following change of the states:
\begin{eqnarray*}
 (\omega_{\tau_1}, \Psi_{\tau_1}) \longrightarrow 
\left( l, \frac{G_l (\tau_1) \Psi_{\tau_1}}
  {\fnorm{G_l (\tau_1) \Psi_{\tau_1}}_{\cH}}\right)
\end{eqnarray*}
\end{itemize}
The algorithm can start again perhaps with other detectors at position (ii).

We want to examine how this algorithm looks like in another reference
frame. Let $\tilde{K}$ be the reference frames which is connected to $K$
by a Poincar\'e-transformation $(\Lambda,a)$ with $\Lambda \in L_+^\uparrow$.

In $\widetilde{K}$, the situation can be described as follows: the particle is prepared at
$\tau_0$ in $\widetilde{x}_0 = \Lambda x_0 + a$ with initial quantum state 
\begin{eqnarray*}
\widetilde{\Psi}_{\tau_0} (\widetilde{x}) & = & (W_{(\Lambda,a)} \Psi_{\tau_0}) (\widetilde{x}) =
S(\Lambda) \Psi_{\tau_0} (\Lambda^{-1}(\widetilde{x}-a))
\end{eqnarray*}
with $\fnormq{\widetilde{\Psi}_{\tau_0}}_{\widetilde{\cH}} = 1$
(because the operator
$W_{(\Lambda,a)}$ is unitary).
The trajectories of the detectors are $\tilde{z}_i  = 
\Lambda z_i + a$, and the detectors are characterized by
\begin{eqnarray*}
\widetilde{G}_j (\tau) = W_{(\Lambda,a)} G_j (\tau) W_{(\Lambda,a)}^+
\end{eqnarray*}
We get $\widetilde{\Lambda}(\tau) = \sum_j \widetilde{G}_j^+
(\tau) \widetilde{G}_j (\tau) = W_{(\Lambda,a)} \Lambda(\tau) W_{(\Lambda,a)}^+$.

Note, that if $\Psi_\tau$ is a solution of \eqref{sec5_dgl} then
$\widetilde{\Psi}_\tau := W_{(\Lambda,a)}  \Psi_\tau$ is a solution of
\begin{eqnarray*}
\fl \frac{\partial}{\partial \tau} \widetilde{\Psi}_\tau
= W_{(\Lambda,a)} \frac{\partial}{\partial \tau} \Psi_\tau
= - \frac{1}{2} W_{(\Lambda,a)} \Lambda(\tau) \Psi_\tau
= - \frac{1}{2} W_{(\Lambda,a)} \Lambda(\tau) W_{(\Lambda,a)}^+
\widetilde{\Psi}_\tau
= - \frac{1}{2} \widetilde{\Lambda}(\tau) \widetilde{\Psi}_\tau
\end{eqnarray*}

This result implies that the algorithm executed in the reference
frame $\tilde{K}$ will give the same detections as the algorithm executed in
$K$ (if we choose the same random numbers). The space-time points of the
detections in the two reference frames are connected by the
Poincar\'e-transformation $(\Lambda,a)$.

The system state $(\omega_\tau, \Psi_\tau)$ in the reference frame $K$ and
the system state $(\widetilde{\omega}_\tau, \widetilde{\Psi}_\tau)$ in the reference frame
$\widetilde{K}$ are always connected in the following way:
\begin{eqnarray*}
(\widetilde{\omega}_\tau, \widetilde{\Psi}_\tau) = 
(\omega_\tau, W_{(\Lambda,a)} \Psi_\tau)
\end{eqnarray*}

The algorithm modelling detections of the particle is indeed covariant.

Now we consider the charge conjugated system $K^C$. Let
\begin{eqnarray*}
\Psi_{\tau_0}^C := C{\gamma^0}^T \Psi_{\tau_0}^* \in \cH^C
\end{eqnarray*}
The charge conjugated coupling is given by
\begin{eqnarray}
G^C_j (\tau) = C {\gamma^0}^T G_j^*(\tau) {\gamma^0}^T C^+
\label{sec5_GC}
\end{eqnarray}
with $G_j^*(\tau)\Psi:=(G_j(\tau)\Psi^*)^*$. Let
\begin{eqnarray*}
\Lambda^C (\tau) = \sum_j G^{C+}_j (\tau) G^C_j (\tau) = 
C {\gamma^0}^T \Lambda^* (\tau) {\gamma^0}^T C^+
\end{eqnarray*}
Note, that if $\Psi_\tau$ is a solution of \eqref{sec5_dgl} then
$\Psi_\tau^C \equiv C {\gamma^0}^T \Psi_\tau^*$ is a solution of
\begin{eqnarray*}
\fl \frac{\partial}{\partial \tau} \Psi^C_\tau
= C {\gamma^0}^T \frac{\partial}{\partial \tau} \Psi^*_\tau
= -\frac{1}{2}C {\gamma^0}^T \Lambda^* (\tau) {\gamma^0}^T C^+ \Psi^*_\tau
= -\frac{1}{2} \Lambda^C(\tau) \Psi^C_\tau
\end{eqnarray*}
We also note that
\begin{eqnarray*}
\fl G^C_j (\tau) \Psi_\tau^C = 
C {\gamma^0}^T G^*_j (\tau) {\gamma^0}^T C^+ C {\gamma^0}^T \Psi_\tau^*  
= C {\gamma^0}^T (G_j (\tau) \Psi_\tau)^* \equiv (G_j (\tau) \Psi_\tau)^C
\end{eqnarray*}
A corollary of this fact is 
$\fnormq{G^C_j (\tau) \Psi_\tau^C}_{\cH^C} =
\fnormq{G_j (\tau) \Psi_\tau}_{\cH}$ and
${<\Psi_\tau^C|\Lambda^C(\tau) \Psi_\tau^C>_{\cH^C}} = 
{<\Psi_\tau | \Lambda (\tau) \Psi_\tau>_{\cH}}$.

We can conclude: if we start with $\Psi_{\tau_0} \in \cH$ and operators $G_j (\tau)$,
then the algorithm will give the same results as if we start with $\Psi^C_{\tau_0} =
C {\gamma^0}^T \Psi_{\tau_0}^* \in \cH^C$ and operators $G_j^C (\tau)$ defined in
\eqref{sec5_GC} (if we choose the same random numbers). 

The state $(\omega_\tau,\Psi_\tau)$  in the normal
system $K$ and the state $(\omega_\tau^C,\Psi_\tau^C)$ in the charge conjugated
system $K^C$ are again connected by
\begin{eqnarray*}
(\omega^C_\tau, \Psi_\tau^C ) = (\omega_\tau, C {\gamma^0}^T \Psi_\tau^*) 
\end{eqnarray*}

Again, we demand that the algorithm applied in the ``charge conjugated
world'' or applied in the ``normal world'' describes the same physical situation. 

In the last part of this section, we examine the non-relativistic limit of the
above algorithm and prove heuristically that the non-relativistic limit reduces to the
algorithm of the non-relativistic EEQT. To establish this fact, we define
\begin{eqnarray*}
 \Omega (\tau, \vec{x}) :=
 (U_\tau \Psi_\tau) (\vec{x})
 = \Psi_{\tau} (c\tau, \vec{x})
\end{eqnarray*}
with $\Psi_\tau$ being a solution of \eqref{sec5_dgl} (we
recall that $U_t \equiv U_{((ct,\vec{0}),\vec{0},\vec{0})}$) and we
assume that $\Psi_\tau \in \hH \; \forall \tau$.
We get
\begin{eqnarray}
\fl \rmi \hbar \frac{\partial}{\partial \tau}\Omega (\tau,\vec{x})
 &= \rmi \hbar c \left( \frac{\partial}{\partial x^0} \Psi_\tau \right)
  (c\tau,\vec{x}) + \rmi \hbar \frac{\partial \Psi_\tau}{\partial\tau}
  (c\tau,\vec{x}) \nonumber \\
\fl &=
  H_D \Psi_\tau (c\tau, \vec{x})
  - \rmi \frac{\hbar}{2} (U_\tau \Lambda(\tau)\Psi_\tau) (\vec{x}) \nonumber \\
\fl &=
  H_D \Omega (\tau, \vec{x})
  - \rmi \frac{\hbar}{2} \left(\sum_j 
  \underbrace{\left[U_\tau G_j^+(\tau)
 U^{-1}_\tau\right]}_{=:g_j^+(\tau)} 
  \underbrace{\left[U_\tau G_j (\tau)
 U^{-1}_\tau\right]}_{=:g_j(\tau)} \right) \Omega (\tau, \vec{x})  
\label{sec5_eq1}
\end{eqnarray}

We examine the non-relativistic limit
of \eqref{sec5_eq1} doing the assumption (in analogy with calculations of the
non-relativistic limit of the Dirac-equation, see e.g. \cite{bjorken.book})
\begin{eqnarray}
\Omega (\tau, \vec{x}) = \fexp{-\rmi \frac{mc^2}{\hbar} \tau} \left(
\begin{array}{c} \phi \\ \chi \end{array} \right)
\label{sec5_eq2}
\end{eqnarray}
Furthermore, we assume that
\begin{eqnarray*}
g_j(\tau) = \left( \begin{array}{cc} g_{j,1} (\tau) & 0 \\
0 & g_{j,2}(\tau) \end{array}
\right)
\end{eqnarray*}
Inserting \eqref{sec5_eq2} in \eqref{sec5_eq1}, we take the
limit $c \to \infty$ but we keep $\frac{e}{c}A^k$. In this way, we
obtain the modified equation of the non-relativistic EEQT (see for
example \cite{blanchard.1995d})
\begin{eqnarray*}
\fl \rmi \hbar \frac{\partial}{\partial \tau}\phi
 & = & \left[ \frac{1}{2m} \sum_l \left( \frac{\hbar}{\rmi} \frac{\partial}{\partial
 x^l} - \frac{e}{c} A^l \right)^2 - \frac{e\hbar}{2mc} \vec{\hat{\sigma}} \vec{B}
 + e A^0 - \rmi \frac{\hbar}{2} \sum_j g_{j,1}^+(\tau)
 g_{j,1}(\tau) \right] \phi
\end{eqnarray*}
with $\hat{\sigma}^k$ being the Pauli-matrices.
We note that
\begin{eqnarray*}
<\Psi_\tau | \Psi_\tau>_{\cH} \, = \int \rmd\vec{x} \, \Omega^+ (\tau,\vec{x})
\Omega (\tau, \vec{x}) \stackrel{c\to\infty}{\longrightarrow} \int \rmd\vec{x} \, \phi^+
(\tau, \vec{x}) \phi (\tau, \vec{x})
\end{eqnarray*}
If we set $t := \tau$, we see immediately that the
algorithm of the EEQT is the non-relativistic limit of the above
relativistic algorithm.

% ===============================================================================
% Section 6
% Summary
% ===============================================================================

\section{Summary}

In this paper, we have presented an alternative version of a relativistic
extension of the Event-Enhanced Quantum Theory (EEQT). It describes
one massive spin $\frac{1}{2}$ particle.

We use the idea of an additional time, the
proper time, which is invariant in all reference frames (in analogy to the
relativistic extension of Blanchard and Jadczyk \cite{blanchard.1996b}). 

The total system consists of a quantum part and a classical part analogously to
EEQT. A pure state $\omega_\tau$ of the classical part at a
proper time $\tau$ is a number ($\omega \in \bN_0$). 
A pure state $\Psi_\tau$ of the quantum part at a proper time $\tau$
is (heuristically) a solution of the Dirac equation. 
We have proved that the solutions of the Dirac equation can
be extended to a separable Hilbert space with a positive-definite scalar
product. An important property of a quantum state $\Psi_\tau$ is that
it is uniquely given by its projection onto a spacelike hyperplane.

The advantage of a positive-definite scalar product must be paid for with a more
complicated Hilbert space compared to the relativistic extension of Blanchard and
Jadczyk \cite{blanchard.1996b}. In that extension, the Hilbert space is simpler
but they use an indefinite scalar product.

The transformation rules of a system state (if we change the reference frame)
have been presented. They are chosen in such a way that the
scalar product between two quantum states is independent of the reference frame.

First, we have postulated a covariant algorithm to simulate ideal, infinitesimal short
measurements. We have shown that the
(non-covariant) standard reduction postulate formulated in a
(preferred) fixed reference frame can be rewritten as a 
special case of our (covariant) algorithm.

Second, we have postulated a covariant algorithm to simulate detections of a
particle. We have shown that the
non-relativistic limit of this algorithm reduces to the PDP algorithm
of the non-relativistic EEQT.

Moreover, we have shown that both algorithms are invariant by charge
conjugation.

We want to end this paper with a summary of the properties of an event
in our theory: 
An event is a change of the (pure) state of the classical part
happened at a proper time. An event can be observed without disturbing
it. We demand
that each event be associated with a point in space-time. In general,
if an event happens, the quantum state changes simultaneously and
instantaneously over the whole space-time.
We do not want to include the principle of
relativistic causality explicitly in our formalism: we even allow
that an event can happen in the past of the previous (concerning
proper time) event. But we want to preserve a weak kind of order, so we demand the
following: no event (e.g. preparation, measurement, or detection) can
be created in a space-time point which lies in the backward light-cone of that
space-time point which is associated with the previously (concerning
proper time) created event. All these demands are fulfilled by the
events generated by our algorithms.

% ===============================================================================
% Acknowledgments
% 
% ===============================================================================

\ack{I would like to thank Ph. Blanchard for many helpful discussions
and the critical reading of the manuscript.
Thanks to A. Jadczyk for some critical comments.}

% ===============================================================================
% References
% 
% ===============================================================================


\section*{References}

\begin{thebibliography}{10}

\bibitem{blanchard.1993a}
Blanchard Ph and Jadczyk A 1993
{\it Phys. Lett.} A {\bf 175} 157-64

\bibitem{blanchard.1995c}
Blanchard Ph and Jadczyk A 1995
{\it Phys. Lett.} A {\bf 203} 260-6

\bibitem{blanchard.1995d}
Blanchard Ph and Jadczyk A 1995
{\it Ann. Phys.} {\bf 4} 583-99

\bibitem{blanchard.2000}
Blanchard Ph, Jadczyk A and Ruschhaupt A 2000
{\it J. Mod. Optic} {\bf 47} 2247-63

\bibitem{aharonov.1984}
Aharonov Y and Albert D Z 1984
{\it Phys. Rev.} D {\bf 29} 228-34

\bibitem{breuer.1998a}
Breuer H P and Petruccione F 1998
{\it J. Phys. A: Math. Gen.} {\bf 31} 33-52

\bibitem{breuer.1998b}
Breuer H P and Petruccione F 1998
{\it Phys. Lett.} A {\bf 242} 205-10

\bibitem{breuer.1999b}
Breuer H P and Petruccione F 1999
{\it Open systems and measurement in relativistic quantum theory}
ed H P Breuer and F Petruccione (Springer)

\bibitem{horwitz.1973}
Horwitz L P and Piron C 1973
{\it Helv. Phys. Acta} {\bf 46} 316-26

\bibitem{fanchi.1993}
Fanchi J R 1993
{\it Found. Phys.} {\bf 23} 487-548

\bibitem{blanchard.1996b}
Blanchard Ph and Jadczyk A 1996
{\it Found. Phys.} {\bf 26} 1669-81

\bibitem{ruschhaupt.1999}
Ruschhaupt A 1999
{\it Preprint} quant-ph/9912060

\bibitem{bjorken.book}
Bjorken J D and Drell S D 1964
{\it Relativistic Quantum Mechanics}
(McGraw-Hill)
\end{thebibliography}
\end{document}